# MODEL CHECKING IN THE COSMA ENVIRONMENT AS A SUPPORT FOR THE DESIGN OF PIPELINED PROCESSING[1]

**Jerzy Mieścicki**[*], **Bogdan Czejdo**[†] **and Wiktor B. Daszczuk**[*]

[*]Institute of Computer Science,
Warsaw University of Technology, ul. Nowowiejska 15/19, PL 00 665 Warsaw, Poland.
e-mail: J.Miescicki@ii.pw.edu.pl , W.Daszczuk@ii.pw.edu.pl

[†] Department of Mathematics and Computer Science,
Loyola University, New Orleans, USA.
e-mail: czejdo@loyno.edu

**Abstract.** *The case study analyzed in the report involves the behavioral specification and verification of a three-stage pipeline consisting of mutually concurrent modules which also compete for a shared resource. The system components are specified in terms of Concurrent State Machines (CSM) and the verification technique used is the temporal model checking in the COSMA environment.*

## 1 INTRODUCTION

In the design of many types of systems (e.g. for measurement, DSP, data acquisition, distributed control) one of often encountered problems is to guarantee the correct cooperation among system components, especially if the system operation involves the use of shared hardware/software resources, like a common memory, buffers between modules, real-time databases, shared hardware units and/or data/control communication paths.

Usually, it is assumed that the given distributed applications run on top of some operating system or other (more or less sophisticated) control mechanism (e.g. DB management system, lower-layer communication protocol, transaction control, bus arbitration unit or even the whole Java virtual machine) which by itself *somehow* manages the problem. This provides the designer with a illusion of safety. However, these mechanisms certainly protect the underlying shared resource against the improper use but do not deal with the proper behavior of applications themselves which call for these resources. Moreover, especially in the case of dedicated real-time systems, the designer often has to implement his/her own low-level system procedures, e.g. for processing interrupts, granting the access to common DMA channels or to real-time databases which (due to the performance requirements) are implemented as RAM arrays rather than as the large relational or OO databases with full-scale transaction control.

This can be a potential source of errors which are extremely hard to discover, identify and correct. Debugging and testing procedures typically applicable to programs do not help much.

---

[1] This work has been supported by grant No.7 T 11 C 013 20 from Polish State Committee for Scientific Research (Komitet Badań Naukowych).

Even the carefully developed functional tests are usually designed to determine whether the system actually *does what it should do*. They do not answer the question if the system eventually *does something it shouldn't* (get deadlocked, for instance). What still worse, the co-operation and synchronization errors are highly nondeterministic by their nature. They occur as a result of the rare sequences or coincidences of events and are hardly predictable. Therefore, neither testing procedures nor the long-lasting simulated or actual execution of a system do not guarantee that these errors would manifest themselves just during the runtime. Apparently, some other approach is needed to protect the system (and its designer) against this type of design errors.

There is a formal procedure that can help, namely the *model checking* [1, 2, 6, 12]. This methodology faces the growing interest in today's computer science and technology. The present paper is aimed to show how model checking techniques, implemented in the COSMA 3.0 software environment, can support the design of concurrent, reactive systems.

The case study for the experiments discussed below came from an industrial real-time data acquisition system called *PowerWatch*, designed and implemented in the Institute of Computer Science, WUT, for the purpose of monitoring and management of power plants. Actually, the system now works 24 hours a day, 7 days a week in five plants throughout Poland. At some stage of system development and upgrade, the decision was made to implement the messaging subsystem (*SigEvent*), schematically shown in Fig. 1. Its purpose is to notify the plant's staff members about events that eventually occur and call for their urgent presence in plant's premises.

Events that need for such an action are detected in an *Event Detector* from the stream of data provided by *SigData*, which is generally responsible for the data acquisition in the whole installation. Then, *Event Filter* decides to whom (and what) a message has to be sent. The *Message Handler* controls the process of sending the messages.

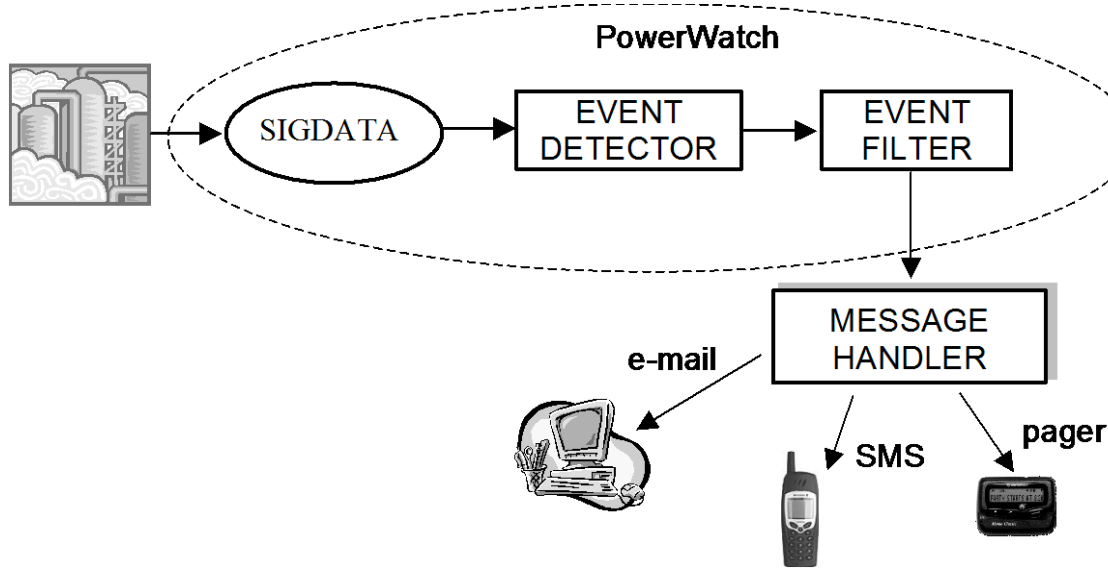

Fig. 1. The idea of messaging subsystem in *PowerWatch*

A closer look on *Message Handler* reveals soon that it should consists of three modules, corresponding to consecutive phases of processing each message is subject to (ordering messages, routing, channel control). So, the *Message Handler* should be designed as pipeline of three modules that provide the smooth and reliable flow of messages from the event filter down to an end user. The number of pipeline stages or modules may vary, but this general functional scheme seems to be typical for a very broad class of applications.

In order to show how model checking methodology developed in the COSMA project can support the design process, the above-mentioned case study was generalized to a system shown in Fig. 2. For the sake of readability, the unnecessary details of *PowerWatch* system are omitted and the functional requirements are simplified so that the specification of components can be illustrated with understandable graphs.

Now, the system under consideration is the pipeline of three processing stages or modules with data source and data sink at two ends of the pipeline. Each module provides the appropriate processing for each data frame it receives from its left-hand neighbor and passes the data to the next (right-hand) one. Additionally, it is assumed that two modules (*Module_1* and *Module_3*) in order to perform their processing, have to use a shared resource, to which the access is mutually exclusive. The functional unit which provides this mutually exclusive access is modeled by the *Arbiter*. To make things still more complex (and realistic), the *Arbiter* deals not only with the two modules of 'our' pipeline, but also with *other* partners (unspecified in more details) which also compete for the common resource.

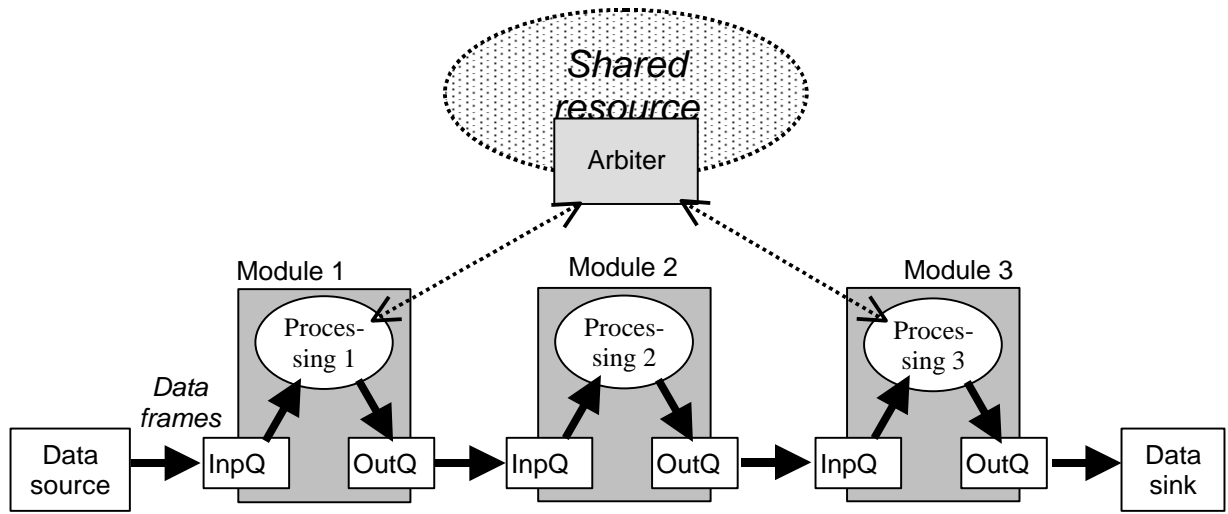

Fig. 2. Flow of data in a three-module pipeline with a shared resource

We show below how these design requirements can be precisely modeled using the Concurrent State Machines (CSM) which are the conceptual framework for the COSMA methodology [27]. The CSM behavioral specification of individual system components is very close to designer's intuition and – to an extent – resembles commonly known UML's state diagrams, statecharts of other finite-state behavioral models. However, in contrast to state diagrams (as well as sequence or cooperation diagrams) CSM are formally precise enough to enable the algorithmic computation of the large graph, showing all system's reachable states and transitions among them. This graph is a global behavioral model of the system and contains *all* its possible executions.  The exhaustive search in the system graph for properties specified as the design requirements is just what the model checking is about. As it can be expected, the main challenge for this methodology is the size of the graph and the complexity of model checking algorithms.

## 2 BRIEF INTRODUCTION TO MODEL CHECKING AND CONCURRENT STATE MACHINES

### 2.1 Model checking

Model checking is a formal method for the verification of systems' behavior. Generally, formal methods for this purpose can be divided into two main groups: theorem proving and finite-state methods [3]. Among the latter ones, model checking enjoys the greatest interest. In the theorem-proving, a question if a given property holds for a given system is expressed as a formal theorem which has to be mathematically proved either true or false. The way how to do it must be devised (almost) each time anew, therefore the verification (even if supported by tools like PVS [15], Larch [17], or HOL [16], see also [9]), requires a sound theoretical background, intuition and formal skills from the designer. It is probably why the theorem proving (however universal and mathematically elegant this approach is) enjoys only a limited interest in practice.

In contrast to theorem proving, model checking offers the designer a still-growing set of ready-to-use algorithms and techniques for the analysis of system's properties. So, it is potentially more acceptable from practitioners' point of view. Generally, the approach consists in the following. First, out of some primary specification of system components (e.g. programs or some program-like notation, Z notation, state diagrams or other finite-state graphs etc.) one has to build a large but *finite* graph containing all possible (reachable) system states and all possible transitions among them. This graph makes the (finite-state) behavioral *model* of a system. Each *path* in the graph represents the allowable execution or the (part of) behavior of a system. The graph contains all possible executions or behaviors. Also, the property in question has to be specified. In a *temporal model checking* the property is specified as a temporal expression, i.e. the formula involving temporal operators ('always', 'eventually', 'until', 'next') in addition to Boolean ones ('and', 'or', 'not') and the quantifiers ('in *all* paths starting from state *x*', 'there *exists* a path starting from *x* such that …'). Then, given the system model and the property, the *exhaustive search* of the system's state graph is performed, aimed to decide whether the desired property holds or not.

As the model (graph) is finite, the problem is decidable. It can be algorithmized so that the designer is offered a repertoire of functions designed for the analysis of numerous aspects of system's behavior. Moreover, if the property does not hold, one can obtain a *counterexample*, i.e. the path leading to the just-identified failure. This provides the extremely valuable feedback information enabling the designer to identify and correct the component which is responsible for a negative outcome of the checking.

The main limitation the model checking is confronted with is the *exponential explosion* of the state space size. Indeed, even for a relatively small system of, say, 10 components (10 states each), a set of all possible states can be of order of $10^{10}$. Even though not all these states are actually reachable (due to synchronization constraints among components), the exponential nature of the system graph size has been always considered a serious limitation for finite state methods. Breakthrough came in early 1980's with the use of Reduced Ordered Binary Decision Diagrams (ROBDD [4, 5]) and the idea of *symbolic model checking* [6, 12]. ROBDD enable the convenient representation of very large graphs while the symbolic model checking allows to reduce the computational effort needed for the evaluation of temporal formulas in a given graph. Soon, model checking became a practical technique for the verification of industrial hardware projects, protocols and software [7, 8, 2]. Nevertheless, the exponential growth of state space is still a real threat. An extensive research is being done on various techniques that can manage the problem.

There are numerous software tools (or *model checkers*) available for these purposes. Among the most frequently referenced ones are SPIN [10, 11], SMV [12, 13], Cospan (or its commercial version, FormalCheck [14]). A few dozen of other tools have been implemented for research purposes [3]. COSMA, the software environment used in this paper, is one of them [18]. It is based on the idea of Concurrent State Machines (CSM), the finite-state model particularly suitable for the modeling the cooperation and communication among the components of concurrent reactive systems, as well as between the system and its environment. Since the beginning of the idea itself ([19, 20]) the CSM approach has been tested in several studies ([21, 22, 23]) and the implementation of the COSMA software tool was undertaken. Its recent upgrade (COSMA 3.0, now under testing) supports structured specification of components and a powerful reduction technique which was also used in a case study discussed in the present paper ([29,30]).

Concurrent State Machines support in a unique way two aspects of concurrency: simultaneous occurrence of communication events (formally – symbols of the input alphabet) and simultaneous execution of component actions. No special mechanism for interleaving actions or sequencing the input is assumed. The system of CSM performs as if it was embedded in a communication medium which instantaneously and faultlessly broadcasts to all system components the union of output symbols produced by the environment and components themselves. However, the delays, nondeterministic loss of symbols, (finite) buffers as well as specific sender – receiver pairs (instead of broadcast-mode communication) can be also modeled, but as a deliberate designer's decision rather than as an implicit general assumption. The advantage of the CSM model is that the *product* of the system (or the large graph representing the possible system behaviors) is relatively simple to obtain (using the state-of-the-art BDD library) and makes an almost ready-for-use Kripke structure needed for the purposes of temporal model checking [25].

Also, the extended version of CSM model (ECSM) [26] is studied and implemented in the COSMA environment. It allows for defining local and global variables and attributing actions (practically – sequential programs) to states and transitions of the finite-state CSM models of components. This supports the design of complex software: first, the correctness of the communication and cooperation among components (viewed as CSM models) is verified by model checking, then - as the 'communication skeleton' of the system is proven correct - the 'real' data structures and actions over them are added. The ultimate goal is to obtain an executable code from the ECSM specification.

## 2.2 Concurrent State Machines

The modeled system $Z$ is a nonempty, finite set of components, written $Z = \{m_1, m_2, …, m_k\}$. Each $m_i$ is a (finite state) Concurrent State Machine or CSM. To define the CSM, one has to specify its graph and indicate one of graph nodes as an initial node. Graphs used for the definition of CSM are referred to below as COSMA Labeled Graphs or CLG.

At the stage of specification, CLG themselves can be viewed as just graphs: collections of nodes, arrows and labels attributed to them. We can edit them, e.g. add or delete nodes and edges or modify the labels attributed to elements of the graph. Also, we can define the more complex operations over CLG graphs: e.g. 'stitching' two unconnected graphs into one graph (by adding some 'stitching edges' with appropriate labels) or, conversely, cutting a sub-graph from a larger graph.

Only once the specification is completed and we choose one of nodes as the initial node - the graph (CLG) becomes the machine (CSM). So, from the same graph we can obtain several different machines, by a different choice of the initial node. Similarly, given the collection of CSM, we can define and analyze many different systems, indicating which machines make the system at the present stage of analysis.

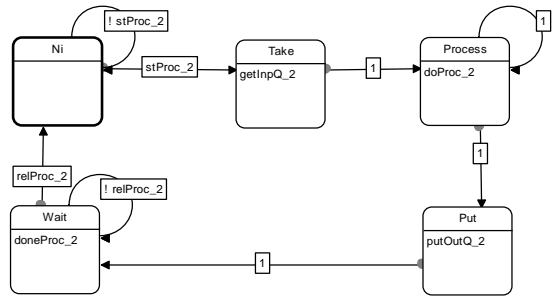

Fig. 3. CSM model of *Proc_2*

Example CSM machine is shown in Fig. 3. It is a model of one of components used later on, namely *Proc_2*. Its CLG graph consists of five nodes (rounded boxes) and a number of directed edges or arrows. One node, namely *Ni* (highlighted with a thicker line) is indicated as machine's initial node. Nodes have their identifiers or names, placed in upper part of the box, e.g. *Ni*, *Take*, *Process* etc. Like in other labeled transition systems, CLG nodes represent states of CSM behavior. At any instant of time, the machine is in exactly one of its states. At the beginning of the observable behavior, machine is in its initial state.

In lower part of the node a set of output symbols (attributed to a given node) is enumerated. The machine transmits these symbols whenever it visits the given node. In the example from Fig. 3, in node *Ni* the machine transmits no symbol (or: empty set of symbols), in *Take* the one-element set {*getInpQ_2*} is transmitted, while in *Wait* it produces {*doneProc_2*}, etc. If a set of two or more output symbols is attributed to a node (which is not the case for this particular machine), then all these symbols are transmitted *simultaneously*, as long as the machine is in the given state. It is assumed that any symbol produced inside the system is produced by only one component, but it can be 'watched to' by multiple communication partners. This way, we can define pairs or small clusters of communicating components, even though in a CSM model (formally) all the symbols are immediately broadcasted to all components.

Directed edges or arrows of a graph are labeled with Boolean formulas. Operators +, *, ! stand for Boolean sum, product and complement (respectively). Generally, the formula specifies the conditions under which the given edge is *enabled*. For instance, formula *stProc_2* (at the arrow from *Ni* to *Take*) means that this transition is enabled if (while the machine is in *Ni*) the symbol *stProc_2* is received by the machine. Similarly, formula !*stProc_2* (at the arrow from *Ni* back to *Ni*) means that this edge is enabled if (in the same state) the symbol *stProc_2* is *not* received, etc. Formula *1* is unconditionally *true*, therefore arrows labeled with it are enabled regardless of machine's input.

If for a given ('present') state two or more outgoing edges are enabled, then *one of them* is selected as *active* edge. This choice is *nondeterministic* and *fair*, so that if the machine visits this particular state infinitely large number of times - each enabled edge is eventually selected also infinitely often. Anyway, the edge chosen as active one points out the next state to be assumed by the machine.

It should be emphasized that in the CSM framework only arcs between two different nodes are referred to as *transitions*. The edges that lead from a node back to the same node ('self-loops' or 'ears', as they are often called in the CSM parlance) are not treated as transitions. They mean that the machine, under a condition specified in the formula attributed to an 'ear', can *remain* in a given state. Indeed, no actual 'transition' is then executed. Thus, for instance, the machine from Fig. 3 remains in *Ni* if *stProc_2* does not come. If *stProc_2* comes - the transition to *Take* is executed. Transitions are instantaneous, i.e. they take no time.

Transitions labeled with *1* are called *spontaneous* transitions. For instance, entering *Take* state the machine produces output symbol *getInpQ_2*, then immediately and spontaneously performs the transition to *Process*. The behavior in *Process* is more complex: it involves both nondeterminism and spontaneity. For this state, two outgoing edges are unconditionally enabled, so the machine faces the nondeterministic choice: to remain in the present state or to perform the transition to *Put*. The above-mentioned fairness condition guarantees that the transition to the next state is eventually (spontaneously) performed. Practically, it means that the machine remains in *Process* for arbitrarily long but finite time. The described mechanism is often used for the representation of the flow of time. In the CSM model (at least by now) there are no means that would provide for the specification of absolute lengths of time intervals: they are introduced only for Extended CSM [26].

To sum up, the example *Proc_2* performs as follows. Initially, it is in its *Ni* state, watching only for starting symbol (signal, message, event) *stProc_2* to come (this event is produced by *Main_2*). When it eventually happens, *Proc_2* spontaneously goes through a sequence of states, where it gets the data from module's input queue, processes it (spending some time in *Process* state), puts the result into module's output queue and notifies the *Main_2* process (issuing the *doneProc_2* event) that it has just performed its job. When released by *Main_2*, it returns back to its initial, idle state and waits to be started again by *Main_2*.

In the CSM framework it is assumed that all the components that make the system are embedded in the common, ideal communication medium which instantaneously (and faultlessly) broadcasts *the set union* of all symbols transmitted by all the components as well as symbols coming to the system from the environment. In each instant of time, the (present) state of system is a vector of present states of components. Each component transmits its output symbols, attributed to its present state. Thus, for any system state, the set of output symbols is the set union of symbols transmitted by all components in their present states. Therefore, in any system state some symbols are received (by all components) while other ones are not: this causes some transitions be enabled etc. However, some system edges certainly can be *never* enabled in a given state and they should be removed from the graph which represents the global system behavior.

In the COSMA framework, the above process of elimination of such 'void' edges, calculating actual formulas and determining actually reachable states is well-defined and algorithmized. The algorithm is described in [27]. Starting from the initial system state, it computes Boolean formulas that label all potential edges outgoing from the given system state and cancels all void edges. The remaining arcs indicate system states which are actually immediately reachable from the given state, then actual formulas are computed etc. The process continues recursively, until no new actually reachable system states are discovered.

This way we obtain the single CSM machine which is the CSM model of joint, actual behavior of whole system. This machine is referred to as *system's reachability graph* or *CSM product* of the set of components. System reachability graph is just the *model* of a system. It makes the Kripke structure [25] necessary for the evaluation of temporal formulas.

## 3   CSM MODEL OF A THREE-MODULE PIPELINE

General functional requirements for the system have been briefly given in Section 1. The system (as shown in Fig. 2) provides unidirectional flow of data frames or messages, so that any message receives a portion of processing in a given module and then it is passed to the next one. Additionally, we have assumed that both *Module_1* and *Module_3*, in order to perform their processing properly, have to compete (with each other and also with additional, unspecified counterparts) for an exclusive access to a shared resource.

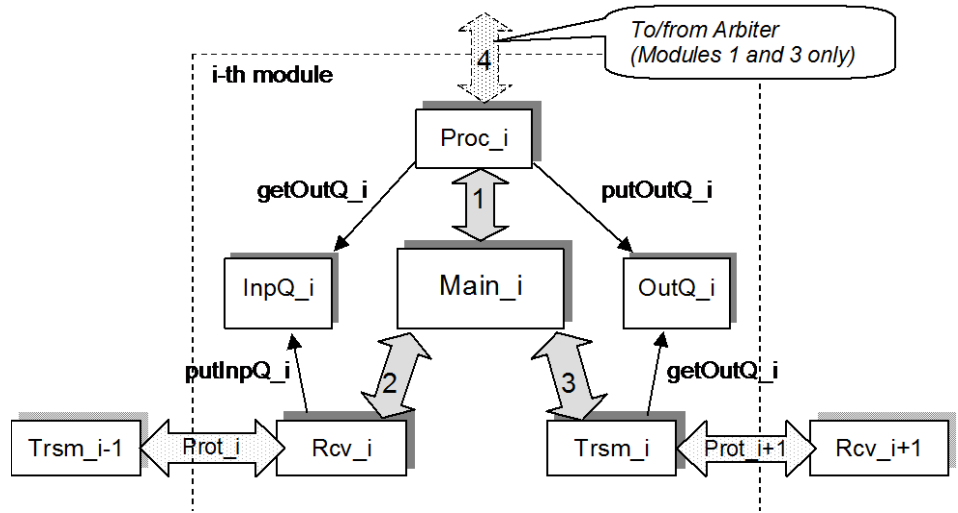

Fig. 4. Structural diagram of a module. 1, 2, 3 – intra-module control interfaces, 4 – interface to *Arbiter* (modules 1 and 3), *Prot_ i, Prot_i+1* – inter-module protocol

The single module is organized as in Fig. 4. While Fig. 2 illustrated schematically the flow of data, Fig. 4 is focused on the module components and on the flow of control among them. The i-th module consists of six following components:
- *Main*, the main process which controls the remaining components,
- *Rcv*, the receiver responsible for obtaining messages from the left-hand neighbor,
- *Trsm*, the transmitter responsible for sending messages to the right-hand neighbor,
- *Proc*, the processing component responsible for *i*-th phase of processing,
- *InpQ*, *OutQ*, input and output buffers of the module.

Arrows in the structural diagram indicate the communication among components. For instance, the arrow from *Rcv* to *InpQ* means that the component *Rcv* transmits (in one of its states) its output symbol *putInpQ* while *InpQ* has (at one of its edges) the Boolean formula referring to the occurrence of the same *putInpQ*. For the sake of readability, some groups of such communication dependences are replaced (in Fig. 4) by two-directional block arrows. The idea is explained in Fig. 5. Interactions between *Main* and *Rcv*, *Main* and *Trsm* as well as between *Main* and *Proc* (indicated in Fig. 4 by arrows numbered 1, 2, and 3) follow the same pattern, shown in the left-hand diagram of Fig. 5. In the case of interaction '2', for instance, *Main* issues *stRcv* that starts the receiver. The *Rcv* has to respond with *doneRcv* as soon as its action terminates. Then, *Main* releases the *Rcv* by transmitting *relRcv*.

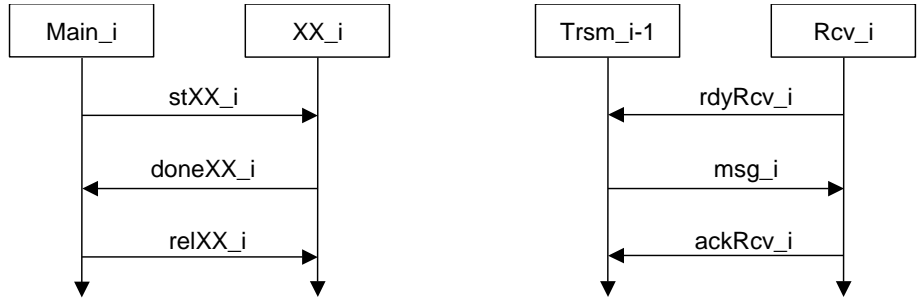

Fig. 5. Sequence diagrams of intra-module control interfaces (left, 'XX' stands for '*Proc*', '*Rcv*' or '*Trsm*'), and inter-module protocol (right)

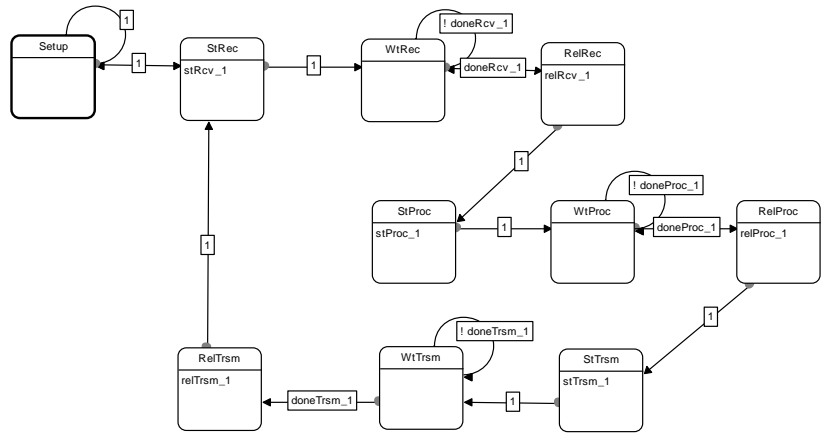

Fig. 6. CSM model of *Main_1*

Interactions between *Trsm_i-1* and *Rcv_i* (inter-module protocol) also involve the exchange of three symbols or events, as shown in the right-hand diagram of Fig. 5. First, *Rcv* issues the symbol *rdyRcv* which notifies the (left-hand neighbor's) transmitter that it is ready to receive. Then, *Trsm* transmits the message *msg* and waits for the acknowledgement *ackRcv*. After *ackRcv* the interaction terminates.

The reader is encouraged to check how the above-mentioned behavioral requirements are modeled by appropriate CSM models of components. The CSM model of *Main_1* is shown in Fig. 6. Apart from the initial *Setup* state, the activity of *Main_1* consists in a cyclically repeated execution of three sections: receiving the message (from *StRec* state up to *RelRec*), processing it (from *StProc* to *RelProc*) and transmitting it to the next module (*StTrsm* to *RelTrsm*).

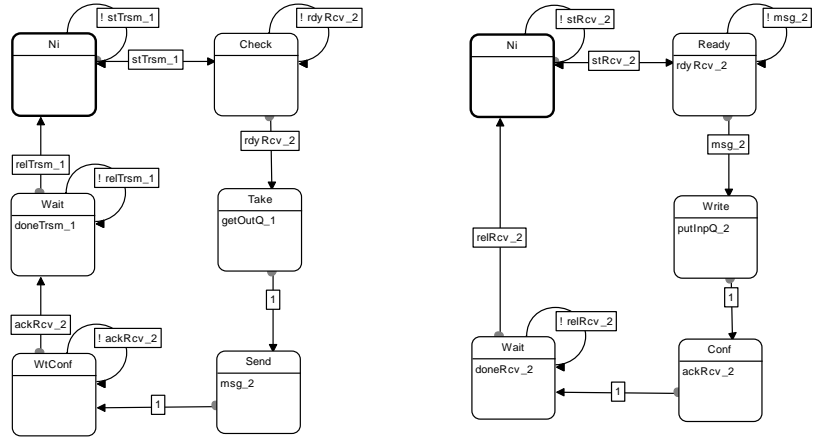

Fig. 7. CSM models of *Trsm_1* (left) and *Rcv_2* (right)

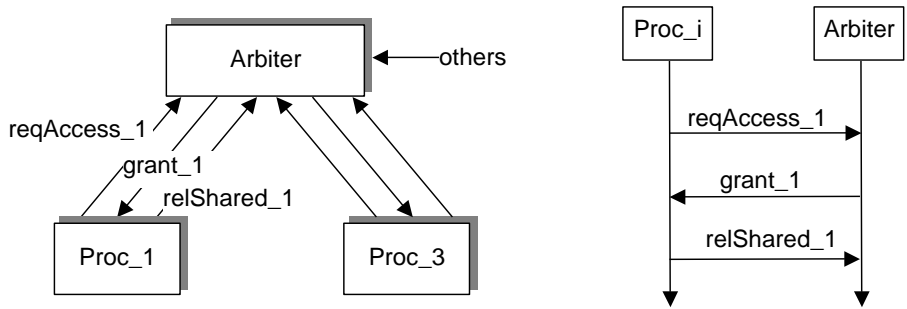

Fig. 8. Interface between *Proc_1* and *Arbiter* (the same for *Proc_3*).

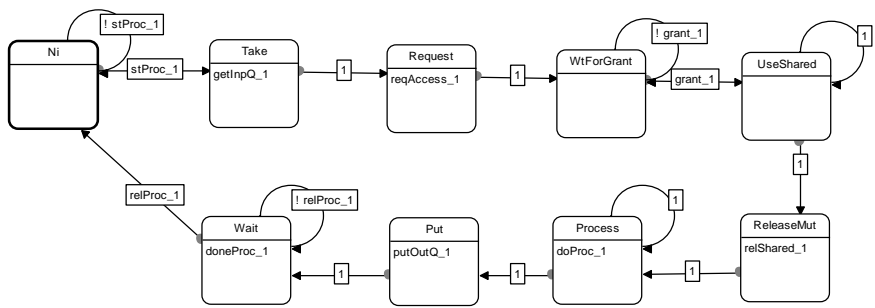

Fig. 9. CSM model of *Proc_1*

Fig. 7 contains the CSM models of *Trsm_1* and *Rcv_2*, so that one can trace the cooperation between *Main_1* (Fig. 6) and *Trsm_1* (intra-module interface #3) as well as between *Trsm_1* and *Rcv_2* themselves (inter-module protocol between modules #1 and #2).

All the components discussed by now (*Main*, *Rcv*, *Trsm*) are identical for all the three modules of the pipeline, of course except for the indices (*_1, _2, _3*) terminating the names of their input and output symbols. However, processing components (*Proc_i*) differ. The CSM model of *Proc_2* was yet shown in Fig. 3. Recall that processing in modules #1 and #3 involves the use of common, shared resource. Therefore, they additionally have to communicate with *Arbiter* to obtain the exclusive access to it. The structural diagram of this part of a model is shown in Fig. 8. and the model of *Proc_1* is in Fig. 9. As it can be expected, *Proc_3* is analogous and is not shown. The Arbiter itself is shown in Fig. 10.

The remaining system components are: input and output buffers, *InpQ* and *OutQ*. (for each module), message source *Trsm_0* and sink *Rcv_4*. Their models are not shown here, the reader is referred to [29,30].

## 4 VERIFICATION OF THE MODEL

Now we have the behavioral model of the same example system (CSM product of 21 components) which has 8284 states and 34711 edges. The model have been verified in COSMA environment, using TempRG module [28]. The system was found incorrect, then the cause of an error was identified and corrected.

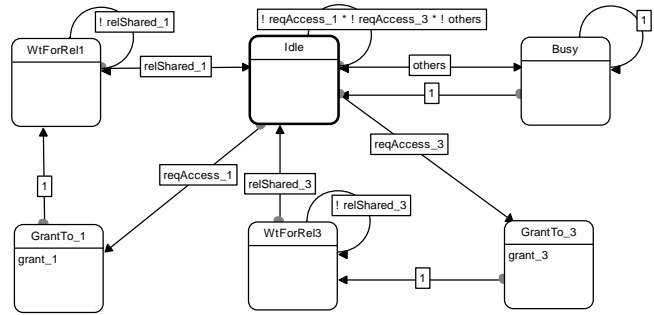

Fig. 10. CSM model of *Arbiter*

To illustrate the methodology, consider the model checking of two types of conditions:
- safety condition (which says, informally, that the system is always in one of 'proper' states)
- liveness condition (that from any of its states the system is always able to reach every of its 'proper' states).

To qualify which system states are 'proper' ones, the following procedure has been applied:
1. A desired *invariant property* of system's behavior was determined,
2. This property was expressed in a form of an additional Concurrent State Machine, called *Invariant*,
3. The new behavioral model has been obtained, combining in one CSM the properties of a system model *and* its invariant,
4. The appropriate temporal formulas have been determined and evaluated in this new model.

The system invariant is determined in terms of number of messages that are currently processed within the system. The system is data-driven pipeline, i.e. the pieces of data may be treated as tokens flowing through the consecutive modules. The capacity of a single module is one token, which follows directly form the construction of a module. So, for the whole system, the number of tokens (messages) processed inside may vary from zero (no messages inside) up to three messages. Each time the *msg_1* is produced by data source – the number of tokens inside is incremented by one. Similarly, each time the *msg_4* is produced (which leaves the system and is annihilated in a data sink) – the number of tokens inside the system is decremented by one. The automaton implementing this property is shown in Fig. 11.

States *s0* through *s3* of *Invariant* correspond to the number of messages inside. As the CSM formalism allows for the simultaneous occurrence of symbols, the Boolean formulas in edges reflect it. For instance, if there are two messages inside (*s2*) and only *msg_4* is produced, then the *Invariant* performs the transition to *s1*, but if *msg_4* occurs simultaneously with *msg_1* – then the number of messages inside is still equal to 2 and the machine remains in *s2*, etc. The *Error* state occurs when the system accepts a new token when three tokens are processed or when a token is retrieved form the empty system. Of course, if the system is correct, *Error* is never reached (safety condition).

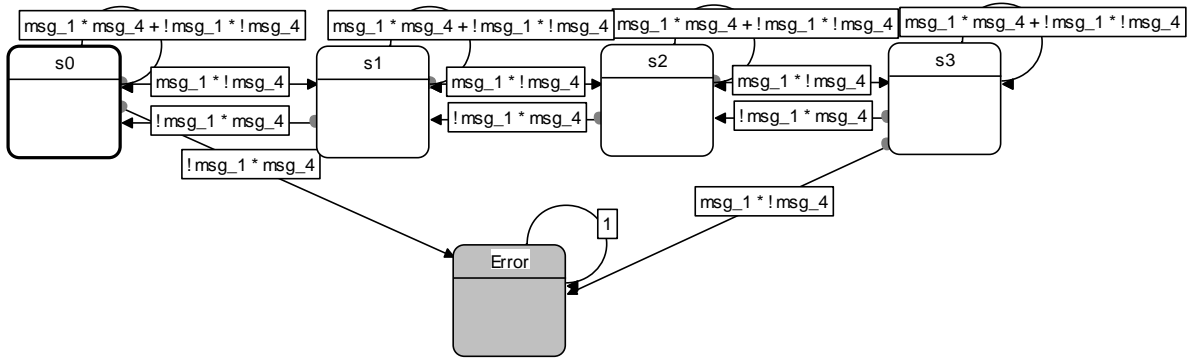

Fig. 11. The *Invariant* automaton

New system's behavioral model, combining the system and the above invariant is a CSM product of {*system*, *Invariant*}, i.e. the machine *system* ⊗ *Invariant*. One can say that the *Invariant* is an additional component, which acts in parallel to the original system itself, 'observes' its behavior and marks each system state with its own state.

The property referred to above as the *safety property* now would say that the new system never visits a state with an *Error* in this element of state vector which corresponds to *Invariant*. This condition is expressed in QsCTL logic [25, 28] as the following temporal formula:

$$\mathbf{AG} \neg \mathbf{in}\, Invariant.Error.$$

The above property was evaluated *true*. So, the system holds it. The evaluation time was 17 seconds on a PC computer with 800MHz processor and 512 MB of RAM.

The *liveness property* says that from any system state, every 'proper' state of *Invariant* (i.e. states *s0* thru *s3*) is reachable. As on CSM systems the fairness condition is imposed [25], the system must actually reach every (non-*Error*) invariant state infinitely often. For our purpose it is sufficient to check if it is true that both 'bounds' of invariant, i.e. states *s0* and *s3*, are reached infinitely often. This requirement is expressed in a form of two following QsCTL formulas:
 1. $\mathbf{AG\, AF\, in}\, Invariant.s0$
 2. $\mathbf{AG\, AF\, in}\, Invariant.s3$

Both these formulas have been evaluated *false* in the model. This negative result means that the system may enter such a state (states) that - from this state on - the pipeline is never empty again (i.e. it never terminates the processing of messages) or is never able to process three messages at once, which it was designed for. The evaluation times were: 54 seconds for formula (1), and 4 min 40 seconds for formula (2).

The model verification can be summarized as follows;
- The system itself performs wrong: there must be a synchronization bug in the specification of components. This calls for the analysis of a counterexample [29,30],
- The advantages of the evaluation algorithm used in the COSMA tool have been also confirmed. The algorithm terminates the evaluation as soon as the result (*true*, *false*) is certainly determined. It is why the evaluation times of rather similar formulas (1) and (2) differ by a few dozen of times.

## 5  CONCLUSIONS

The general message of the present paper is that the COSMA-style model checking is a valuable methodology that can substantially support the design process, especially at its early stage. The system viewed as a collection of cooperating CSM components can be treated as system's preliminary behavioral specification which provides the reasonable insight into the cooperation of modules. Moreover, the model checking procedures can identify the possible synchronization pitfalls and provide the clues when such an error may occur and how one can modify the design to prevent it. Once this is done, the CSM specification of individual components can be a basis for the further refinement (also using state diagrams, sequence diagrams and other means supported by professional CASE tools) but the yet-identified (and corrected) errors do not propagate into the consecutive design stages.